\documentclass[aps,reprint,twocolumn,showkeys,11pt, notitlepage,nofootinbib]{revtex4-1}
\usepackage[colorlinks,allcolors=blue]{hyperref}
\usepackage{graphicx,epsfig}
\usepackage{subfigure}
\usepackage[squaren]{SIunits}
\usepackage{amsmath}
\usepackage{amsfonts}
\usepackage{amssymb}
\usepackage{longtable}
\usepackage{cancel}
\usepackage{booktabs}
\usepackage{tcolorbox}
\usepackage{threeparttable}
\usepackage{subfigure}
\newtcbox{\mybox}[1]{nobeforeafter,colframe=#1,colback=white,boxrule=0.5pt,arc=2pt,
  boxsep=0pt,left=3pt,right=3pt,top=3pt,bottom=3pt,tcbox raise base}

\begin{document}
{
\title{Generation of High-Power, Tunable Terahertz Radiation from Laser Interaction with a Relativistic Electron Beam }
\author{Zhen \surname{Zhang}}
\email{zhen-zhang12@mails.tsinghua.edu.cn}
\author{Lixin \surname{Yan}}
\author{Yingchao \surname{Du}}
\author{Wenhui \surname{Huang}}
\author{Chuanxiang \surname{Tang}}
\affiliation{Department of Engineering Physics, Tsinghua University, Beijing 100084, China}
\author{Zhirong \surname{Huang} }
\email{zrh@slac.stanford.edu}
\affiliation{SLAC National Accelerator Laboratory, Menlo Park, CA 94025, USA}
\date{\today }

\begin{abstract}
We propose a method based on the slice energy spread modulation to generate strong subpicoseond density bunching in high-intensity relativistic electron beams. A laser pulse with periodic intensity envelope is used to modulate the slice energy spread of the electron beam, which can then be converted into density modulation after a dispersive section. It is found that the double-horn slice energy distribution of the electron beam induced by the laser modulation is very effective to increase the density bunching. Since the modulation is performed on a relativistic electron beam, the process does not suffer from strong space charge force or coupling between phase spaces, so that it is straightforward to preserve the beam quality for further applications, such as terahertz (THz) radiation and resonant excitation of plasma wakefield. We show in both theory and simulations that the tunable radiation from the beam can cover the frequency range of 1\,$\sim$\,10\,THz with high power and narrow-band spectra.
\end{abstract}
\maketitle

\section{Introduction}
High-brightness electron beams have been used to drive free-electron lasers (FELs)\,\cite{FLASH,LCLS,SCALA}, high-intensity terahertz (THz) radiation\,\cite{THz1,THz2}, advanced accelerators\,\cite{plasma1,plasma2,dielectric} and beyond. The precise manipulation of the beam phase-space distribution is often desired to advance the development in the above mentioned fields. Recently, there has been great interest in the generation and control of high frequency structure in the current profile of relativistic electron beams. The bunched beam at picosecond (ps) and sub-ps scale can be used to produce intense narrow-band THz radiation\,\cite{THz2,THz3} and resonant wakefield excitation\,\cite{bunch_train1,bunch_train2,bunch_train3,bunch_train4}. Besides, In FELs, sub-ps bunched beams have been applied to generate multicolor X-rays\,\cite{multi-color} based on the sideband effect\,\cite{sideband}.

There are several methods proposed and studied for the generation of ps and sub-ps bunching in electron beams, including exchanging transverse modulation to longitudinal distribution\,\cite{exchange1,exchange2}, direct modulating the drive laser\,\cite{laser_modulation1,laser_modulation2,laser_modulation3,laser_modulation4} and converting wakefield induced energy modulation to density bunching\,\cite{energy_modulation1,energy_modulation2,energy_modulation3}. As the development of the state-of-the-art laser technologies, laser-based manipulation of the electron beam has been implemented widely. An external-injected laser pulse which co-propagates with the electron beam inside an undulator can create energy modulation on the scale of the laser wavelength. The energy modulation can be converted into density modulation by letting the beam pass through a longitudinally dispersive element, e.g., a magnetic chicane. However, the typical laser wavelength is usually $\sim$\,800\,nm, which is much shorter than the THz wavelength. In this case, the concept of difference frequency can be used to approach THz range with two wavelengths of energy modulation in two separated undulator sections\,\cite{diff_frequency1,diff_frequency2,diff_frequency3}.

For the laser-based modulation method, since it is performed directly on a relativistic electron beam, the process will not suffer from the strong space charge force or coupling between transverse and longitudinal phase spaces, which is critical to preserve the beam quality for further applications. For example, in resonant wakefield excitation, the transverse focus size of the electron beam needs to match the plasma density\,\cite{bunch_train1}, which puts a challenging constraint on the transverse normalized emittance.

In this paper, we propose a method based on the slice energy spread modulation (SESM), rather than energy modulation, to generate sub-ps density bunching in a relativistic electron beam. The slice energy spread of the beam will be modulated in a laser heater\,\cite{laser_heater1,laser_heater0} if the laser intensity envelop varies periodically. The device has been used widely in the FEL facilities\,\cite{laser_heater2,laser_heater3,laser_heater4,laser_heater5}. Then after a dispersive section, strong density bunching can be obtained whose wavelength depends on the scale of laser envelope variation, not the laser wavelength. Similar methods have been adopted to generate THz radiation in a storage ring\,\cite{THz2} and multicolor X-rays\,\cite{multi-color}. In this paper, we present detailed theoretical analysis on the conversion from SESM to density bunching and compare it with simulation results. We will also provide an example to produce tunable intense narrow-band THz radiation ranging from 1 to 10\,THz. This method is well suited with the configuration of linear accelerator facilities and is of great potential in applications.

This paper is organized as follows. In Sec.\,\ref{theory}, detailed theoretical analysis on the converstion from SESM to density bunching is presented. It is found that the double-horn slice energy distribution from the laser modulation will help increase the available bunching factor. In Sec.\,\ref{THz}, we propose a dedicated beam line to generate intense narrow-band THz radiation based on the SESM method. Simulations are performed to demonstrate the tunable bunching frequency. Some discussions about the advantages and requirements of the proposed method will be presented in Sec.\,\ref{discussion}. Lastly, we will give a short summary in Sec.\,\ref{summary}.

\section{Theoretical analysis}\label{theory}
Let us assume the initial beam is uniform in current, but has a Gaussian slice energy spread $\sigma_\delta(z_0)$ that is a function of the longitudinal coordinate $z_0$ within the beam. The distribution can be written as
\begin{align}
f_0(\delta_0,z_0)=\frac{I_0}{\sqrt{2\pi}\sigma_\delta(z_0)}\exp\left[-\frac{\delta_0^2}{2\sigma_\delta(z_0)^2}\right]\,,
\end{align}
where $\delta_0$ is the relative energy deviation and $I_0$ is the beam current. Due to the laser modulation, the slice energy spread is assumed to be
\begin{align}
\sigma_\delta(z_0) = \bar{\sigma}\left[1+A\sin(k_0z_0)\right]\,,
\end{align}
where $\bar{\sigma}$ is the average rms slice energy spread $0<A<1$ is the relative modulation depth. The longitudinal phase space is not uniform but integration over $\delta_0$ yields uniform current $I_0$.

An energy chirp is added in the following linac section by $\delta=\delta_0+hz_0$, and then the longitudinal dispersion occurs in a magnetic chicane as
\begin{align}
z = z_0+R_{56}\delta=z_0+R_{56}(\delta_0+hz_0)\,.
\end{align}
Here we ignore the scale of relative energy spread due to the change of beam average energy, which can be absorbed into other parameters. The density modulation appears after the chicane with the bunching factor as follows
\begin{align}
b(k)=\frac{1}{I_0}\int d\delta dz e^{-ikz}f(\delta,z)\,.
\end{align}
Using the Liouville theorem that $f(\delta,z)=f_0(\delta_0,z_0)$, $d\delta dz=d\delta_0dz_0$ and making a change of variable from $\delta_0$ to $\eta = \delta_0/\left[1+A\sin(k_0z_0)\right]$, we obtain
\begin{widetext}
\begin{align}
b(k) & =\int d\delta_0dz_0 e^{-ik(1+hR_{56})z_0-ikR_{56}\delta_0}\frac{1}{\sqrt{2\pi}\sigma_\delta(z_0)}\exp\left[-\frac{\delta^2}{2\sigma_\delta(z)^2}\right]\nonumber\\
& = \int d\eta dz_0e^{-ik(1+hR_{56})z_0-ikR_{56}\eta\left[1+A\sin(k_0z_0)\right]}\frac{1}{\sqrt{2\pi}\bar{\sigma}}\exp\left[-\frac{\eta^2}{2\bar{\sigma}^2}\right]\nonumber\\
& = \frac{1}{\sqrt{2\pi}\bar{\sigma}}\int d\eta\exp\left[-ikR_{56}\eta-\frac{\eta^2}{2\bar{\sigma}^2}\right]\int dz_0e^{-ik(1+hR_{56})z_0}\sum_nJ_n(kR_{56}\eta A)e^{-ink_0z_0}\,.
\end{align}
\end{widetext}
Integration over $z_0$ yields nonvanishing bunching at the wavenumber $k_n = nk_0/(1+hR_{56})$ with the bunching factor
\begin{align}\label{bunching}
b_n = \frac{(-1)^n}{\sqrt{2\pi}\bar{\sigma}}\int d\eta J_n(k_nR_{56}A\eta)e^{-ik_nR_{56}\eta-\frac{\eta^2}{2\bar{\sigma}^2}}\,.
\end{align}
Numerical calculation of Eq.\,\eqref{bunching} can be used to find the exact bunching factor and current distribution. 

Here we focus only on the fundamental modulation wavenumber $k_1=k_0/(1+hR_{56})$, the bunching factor $b_1$ is plotted with a solid line for different $k_1R_{56}\sigma_0$ in Fig.\,\ref{bunching_fundamental}.  The expression of $b_1$ can be simplified if we approximate $J_1(x)\approx a\sin(bx)$ for $|x|<3$ with $a=0.58,b=0.85$, then
\begin{align}\label{approximation}
b_1=\frac{a}{2}[e^{-\frac{k_1^2R_{56}^2\bar{\sigma}^2(1-bA)^2}{2}}-e^{-\frac{k_1^2R_{56}^2\bar{\sigma}^2(1+bA)^2}{2}}]\,.
\end{align}
The optimal chicane setting to achieve the largest bunching factor is to satisfy $|k_1R_{56}\bar{\sigma}|\approx\sqrt{2+0.5A^2}$, and the maximum bunching factor is $\sim$\,0.27.

\begin{figure}[ht]
   \centering
   \includegraphics*[width=80mm]{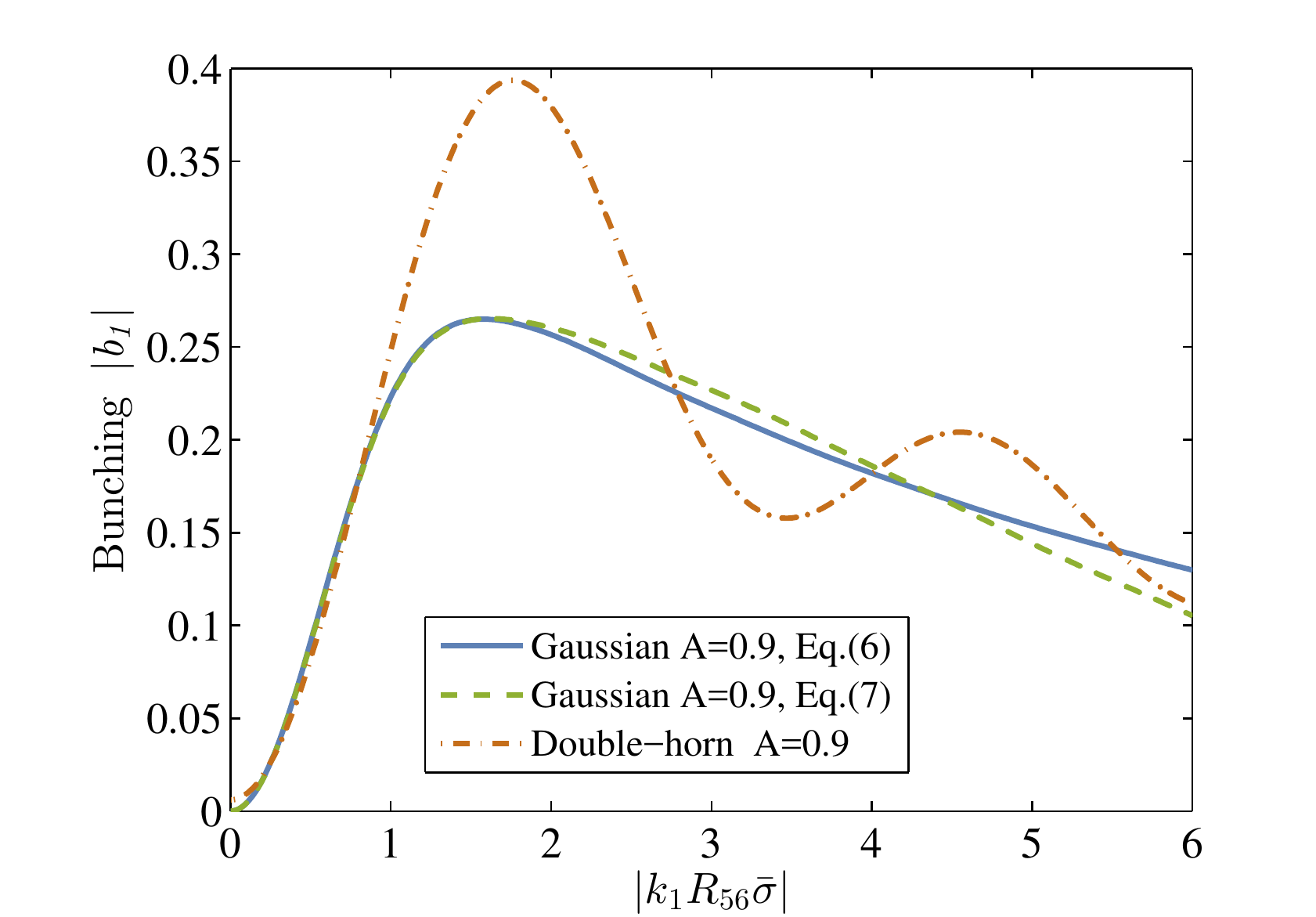}
   \caption{Bunching factor evolution versus $k_1R_{56}\bar{\sigma}$ with $A=0.9$ and different slice energy distributions: (solid line) Gaussian distribution, (dashed line) Gaussian distribution but the bunching factor is calculated by Eq.\,\ref{approximation} and (dotted line) double-horn distribution from laser heater. }
   \label{bunching_fundamental}
\end{figure}

It should be noted that there is an assumption for the above derivations that the beam has a Gaussian slice energy distribution. This assumption, however, is not always true if we use a laser heater to modulate the beam's slice energy spread. In Ref.\,\cite{laser_heater1} it is found that when the laser size is much larger than the electron beam size, the resulting energy profile is a double-horn distribution, which will not contribute much to suppress the instability. Figure\,\ref{double_horn} shows an example of double-horn distribution of the beam slice energy. The detailed parameter settings can be seen in Table\,\ref{parameter} in the following sections. However, with the purpose to increase the density bunching, we find that the double-horn energy distribution is more effective to increase the bunching factor in this study.

\begin{figure}[ht]
   \centering
   \includegraphics*[width=80mm]{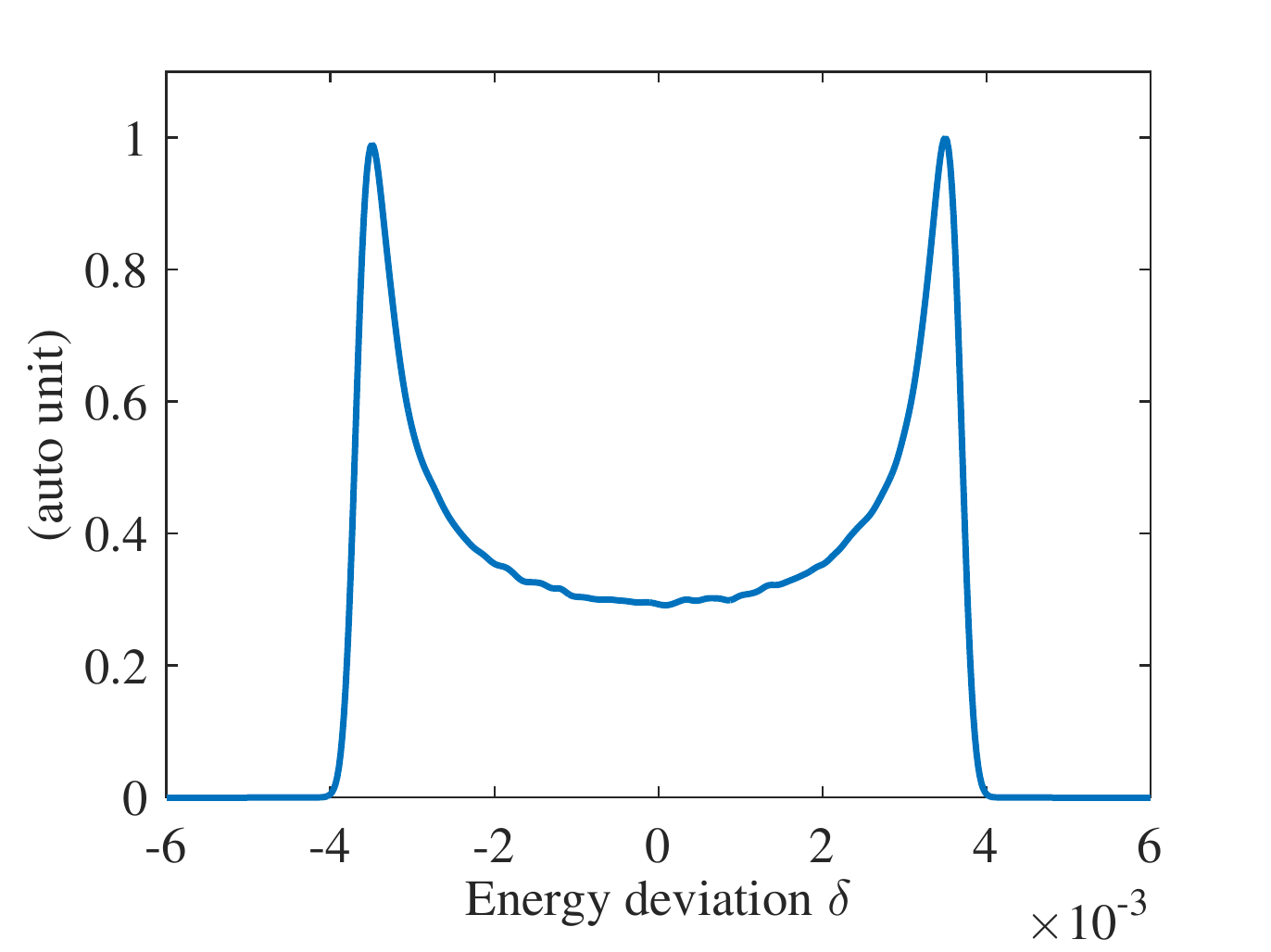}
   \caption{Double-horn slice energy distribution of the electron beam when the laser spot size is much larger than the beam size in the laser heater. The detailed parameters can be found in Table\,\ref{parameter}. }
   \label{double_horn}
\end{figure}

Assuming the laser spot size is larger than the beam size and the laser power profile has the form $\sqrt{P_L(s)}=\sqrt{\bar{P}}(1+B\cos(k_0s))$, the energy modulation in the laser heater can be written as
\begin{align}
\delta_f = \delta_i + \Delta(1+B\cos(k_0s))\,,
\end{align}
with initial and final relative energy deviation $\delta_i,\delta_f$ and energy modulation amplitude $\Delta$. The full expression for $\Delta$ can be found, e.g. in Ref.\,\cite{laser_heater1}. To compare withe Gaussian distribution, we define an effective modulation depth
\begin{small}
\begin{align}
A &= \frac{\sqrt{\Delta^2(1+B)^2+\sigma_0^2}-\sqrt{\Delta^2(1-B)^2+\sigma_0^2}}{\sqrt{\Delta^2(1+B)^2+\sigma_0^2}+\sqrt{\Delta^2(1-B)^2+\sigma_0^2}}\nonumber\\
&\approx 1-\frac{2\sqrt{\Delta^2(1-B)^2+\sigma_0^2}}{\Delta(1+B)}\,,
\end{align}
\end{small}
where $\sigma_0$ is the initial rms energy spread. When $\Delta/\sigma_0=10,B=1$, the modulation depth $A=0.9$. Numerical calculation results of the bunching factor for different $R_{56}$ are presented in Fig.\,\ref{bunching_fundamental}. The optimal $R_{56}$ is similar with the Gaussian case, but the maximum bunching factor is increased to $\sim$\,0.4 with the optimal condition $|k_1R_{56}\bar{\sigma}|\approx 1.75$.

\begin{figure}[ht]
   \centering
   \includegraphics*[width=80mm]{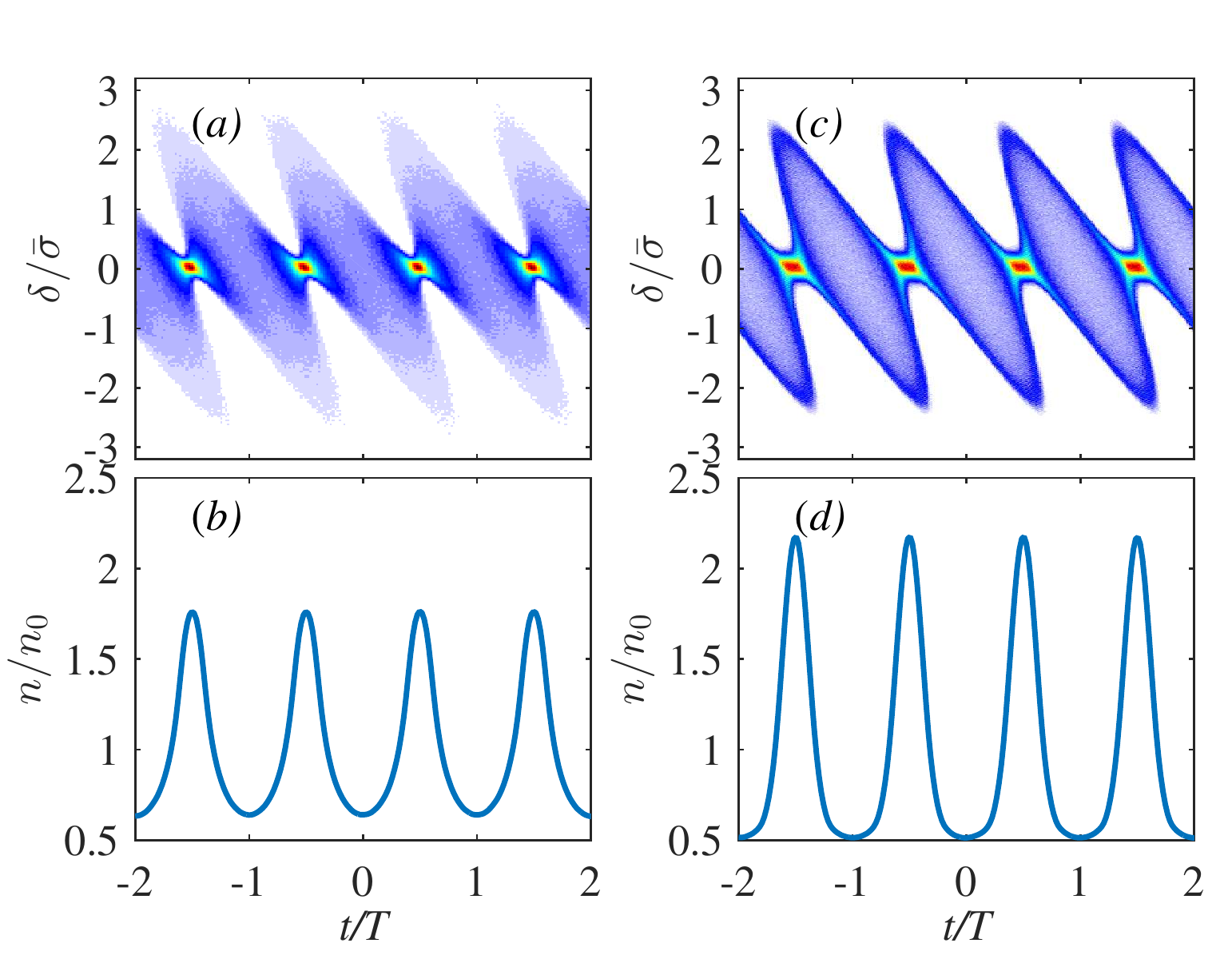}
   \caption{Longitudinal phase spaces and the corresponding density profile with the largest bunching factor in the two different energy distributions with the same modulation depth $A=0.9$. The energy is scaled by the rms energy spread and the time is scaled by the period of the bunching. ($a$) and ($b$) are Gaussian energy distribution and ($c$) and ($d$) are double-horn distribution. }
   \label{phase_space}
\end{figure}

The longitudinal phase spaces and the corresponding density profiles of the electron beam that give the largest bunching factor in both cases are presented in Fig.\,\ref{phase_space}. The physical meaning of $k_1R_{56}\bar{\sigma}$ denotes the rotation angle of the phase space. As the increase of the $R_{56}$, the phase space begins to rotate counter-clockwise and the projected density profile will have peaks and valleys. At a certain angle, the bunching factor reaches its largest value. The similar rotation angle of the two phase spaces in Fig.\,\ref{phase_space} verifies that they both achieve optimal bunching at similar $k_1R_{56}\bar{\sigma}$. In order to optimize the bunching for different parts of the beam at the same $R_{56}$, the modulation depth and the average slice energy spread needs to be uniform over the whole beam. In the double-horn distribution case, the peaks in density profile are higher and sharper, resulting in a larger bunching factor.

\section{THz radiation generation}\label{THz}
From the theoretical analysis in the previous section, the proposed method using SESM can produce strong density bunching in an relativistic electron beam with a bunching factor about 0.4. Based on this method, we propose a dedicated beam line design for tunable intense narrow-band THz radiation generation in Fig.\,\ref{THz_beamline}, which can be implemented in most x-ray FEL facilities or dedicated small THz facilities. The electron beam is generated and accelerated by the gun and the first section of linac before it goes into the modulator. In a typical laser heater design, the undulator is located in the center of a chicane to smear out the laser-induced energy modulation. However, we do not have to do this since the density bunching scale of interest here is much larger than the laser wavelength. The drive laser of the gun and the modulation laser can share the same laser system to make the synchronization between electron beam and laser easier. After the modulator, the beam will pass through the second section of linac with -/+90\,$^\circ$ off-crest acceleration phase to add the energy chirp for THz frequency control without net energy gain. The final chicane is set at optimal value to convert the SESM to density modulation and the radiation frequency is varied by tuning the energy chirp from the second linac. In practice, the average slice energy spread $\bar{\sigma}$ can be calculated after removing the existing energy chirp.

\begin{figure}[ht]
   \centering
   \includegraphics*[width=80mm]{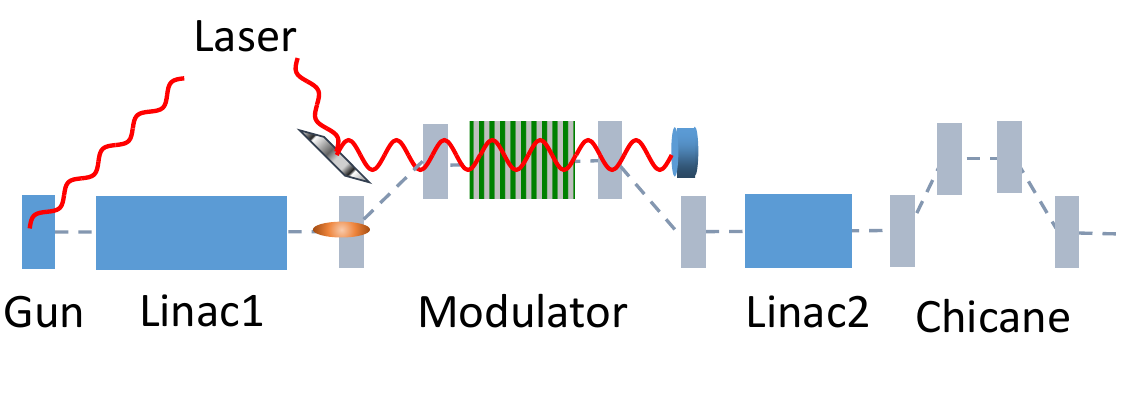}
   \caption{Schematic layout of the proposed beam line for tunable intense narrow-band THz radiation generation. }
   \label{THz_beamline}
\end{figure}

The laser pulse with oscillating power envelop can be generated by chirped pulse beating\,\cite{chirped_beating} or pulse stacking techniques. In the previous work\,\cite{THz2,multi-color}, the laser are both modulated by the first method. Here we discuss the pulse stacking method with $\alpha$-BBO birefringent crystals\,\cite{BBO1,BBO2,BBO3}. Taking 2\,THz density modulation as an example, we need to stack laser pulse with a uniform separation 0.5\,ps. Using crystals with decreasing thickness and assuming the initial laser pules is of Gaussian profile with rms width 60\,fs, the numerical simulation of the stacked laser pulse train is shown in Fig.\,\ref{laser_pulse}. For comparison, we also plot the sinusoidal modulation of the same period with full modulation depth (the intensity minimum along the laser pulse is zero).  

\begin{figure}[ht]
   \centering
   \includegraphics*[width=80mm]{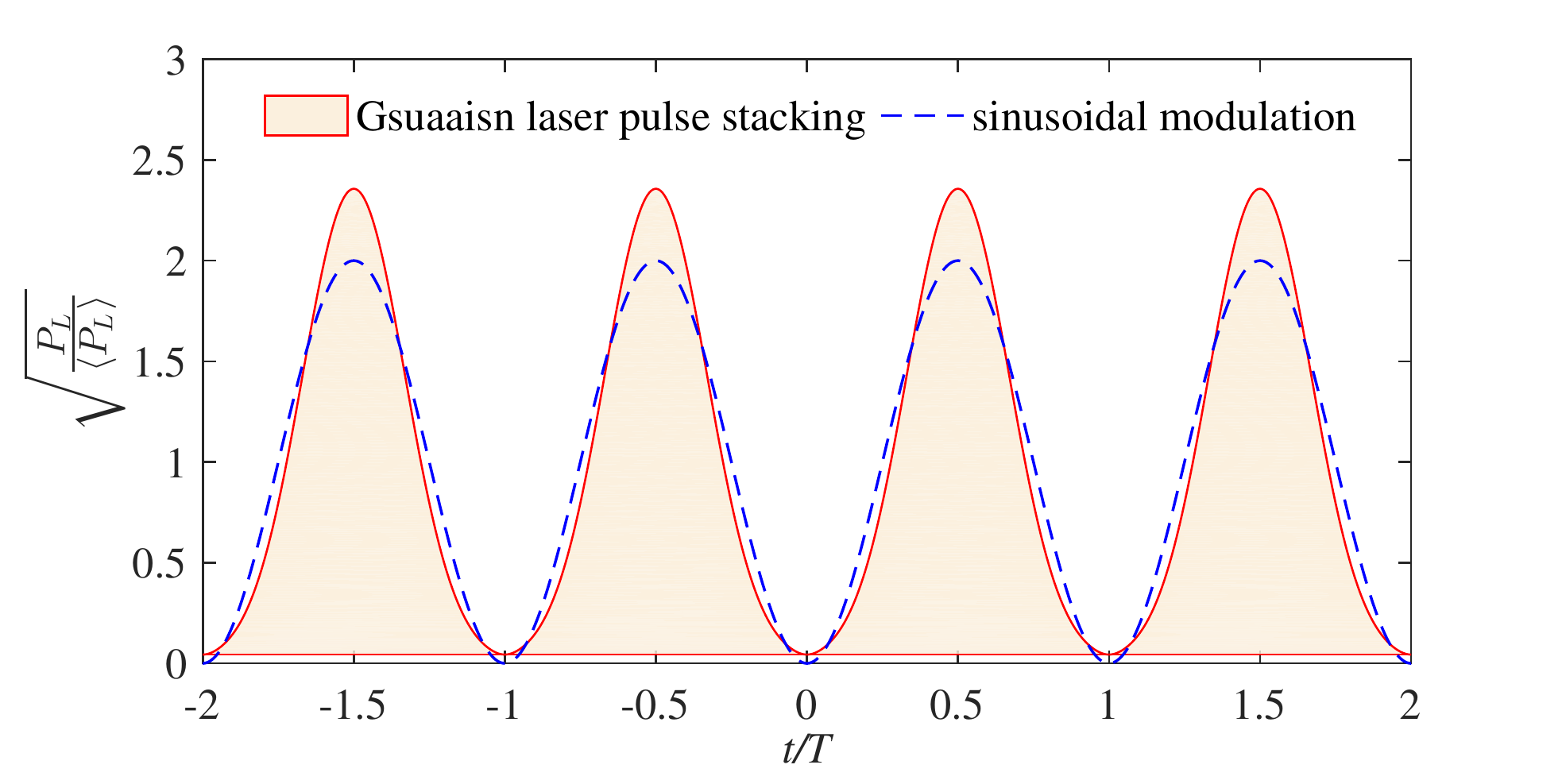}
   \caption{Laser pulse train distribution. The dashed line denotes the full sinusoidal modulation profile for comparison. }
   \label{laser_pulse}
\end{figure}

We use the particle tracking code Elegant\,\cite{elegant} to simulate the laser modulation and beam dynamics. For similarity, the simulation starts from the end of the first linac with the beam and modulator parameters listed in Table\,\ref{parameter}, which is realistic and readily available from the injector of many FEL facilities. The current profile of the electron beam is $\sim$\,10\,ps flop-top distribution. The laser envelop modulation period is 2 and 4\,THz, respectively. Higher frequency modulation can be generated by the chirped pulse beating technique. The waist size of the laser in the modulator is 1.5\,mm, much larger than the electron beam size to generate double-horn slice energy distribution as discussed in the previous section. 

\begin{table}[h!tbp]
\caption{\label{parameter}Beam parameters before the modulator in the simulations.}
\begin{ruledtabular}
\begin{tabular}{c c c}
Parameter&Value&Units\\
\hline
\multicolumn{3}{c}{Electron beam} \\
\hline
Charge & 500 & pC\\
Beam energy  & 135 & MeV\\
Current Profile & flat-top & /\\
Bunch length & $\sim$\,10 & ps\\
Intrinsic slice energy spread & $10^{-4}$ &/ \\
Norm. emittance & 1 & mm-mrad\\
rms beam size & 200 & $\mu$m\\
\hline
\multicolumn{3}{c}{Modulator}\\
\hline
Laser wavelength & 800 & nm\\
Undulator period & 5 & cm\\
Period number & 10 & / \\
Laser waist size & 1.5 & mm \\
Laser stacking separation & 0.5 (0.25$^*$) &ps\\
rms laser pulse length & 60 (30) & fs\\
Laser power & 1 (0.26) & GW
\end{tabular}
\begin{tablenotes}
\item[*] * The numbers in brackets are the parameters for 4\,THz case.
\end{tablenotes}
\end{ruledtabular}
\end{table}

As obtained from the theory, the bunching factor optimizes when $|k_1R_{56}\bar{\sigma}\approx 1.75|$. For one certain frequency of density bunching, we need to match the $R_{56}$ and the average slice energy spread after the modulator. The energy spread is controlled by the laser power. For the 2\,THz case, the laser peak power is 1\,GW and the average energy spread after modulator is 190\,keV (1.4e-3), leading to the optimal $R_{56}=-29.4$\,mm. To verify the theoretical prediction, we turn off the Linac2 and scan the $R_{56}$ of the chicane in the simulation. The bunching factor and the peak current are presented in Fig\,\ref{chicane_2THz}. The optimal $R_{56}$ with the largest bunching factor is -29\,mm, which agrees well with the theory. We also present the longitudinal phase space of the beam at the optimal chicane settings in Fig.\,\ref{phasespace}. The residual energy chirp on the beam is due to the linac wakefield and longitudinal space charge effects.

\begin{figure}[ht]
   \centering
   \includegraphics*[width=80mm]{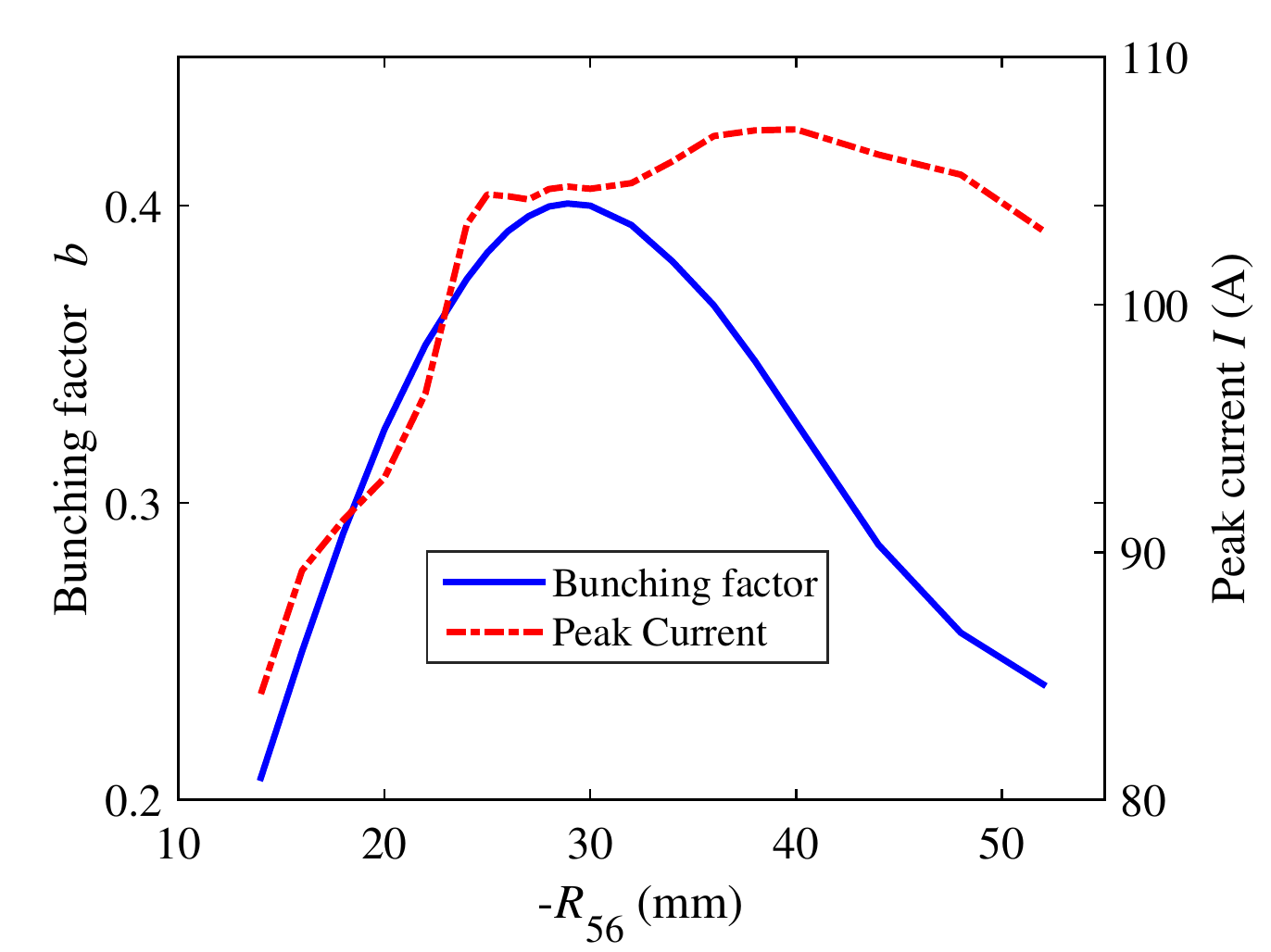}
   \caption{Bunching factor and peak current of the electron beam with fixed laser power 1\,GW and zero energy chirp added onto the beam. }
   \label{chicane_2THz}
\end{figure}

\begin{figure}[ht]
   \centering
   \includegraphics*[width=80mm]{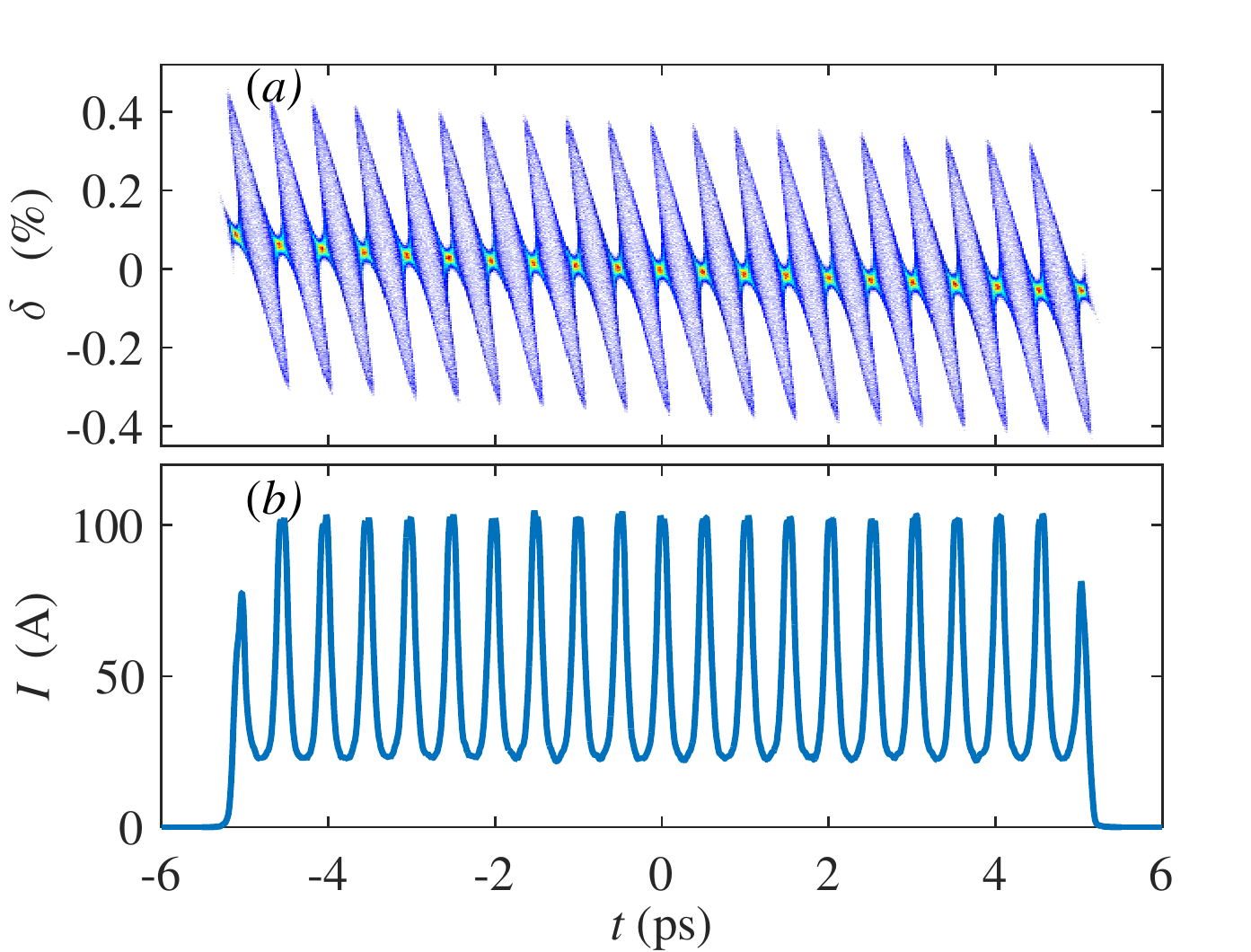}
   \caption{Phase space of the electron beam with largest bunching factor. $R_{56}=-29$\,mm and laser power 1\,GW. }
   \label{phasespace}
\end{figure}

The bunching frequency can be controlled by the induced energy chirp from the Linac2 (an S-band acceleration structure in the simulation) before the chicane. As the bunching frequency changes, we need to vary $R_{56}$ and laser power to maintain the bunching factor. If we fix the laser power at 1\,GW, the $R_{56}$ needs to be varied as shown in Fig.\,\ref{scan_R56}. The bunching frequency can be varied from 1 to 3\,THz with almost constant bunching factor 0.4. We also show the full width of half maximum (FWHM) of the bunching spectra and the requirement for the chirp energy of Linac2. The degradation of the bunching factor and the increase of the spectra bandwidth are both due to the nonlinear effects during the beam compression or stretching.

\begin{figure}[ht]
   \centering
   \includegraphics*[width=80mm]{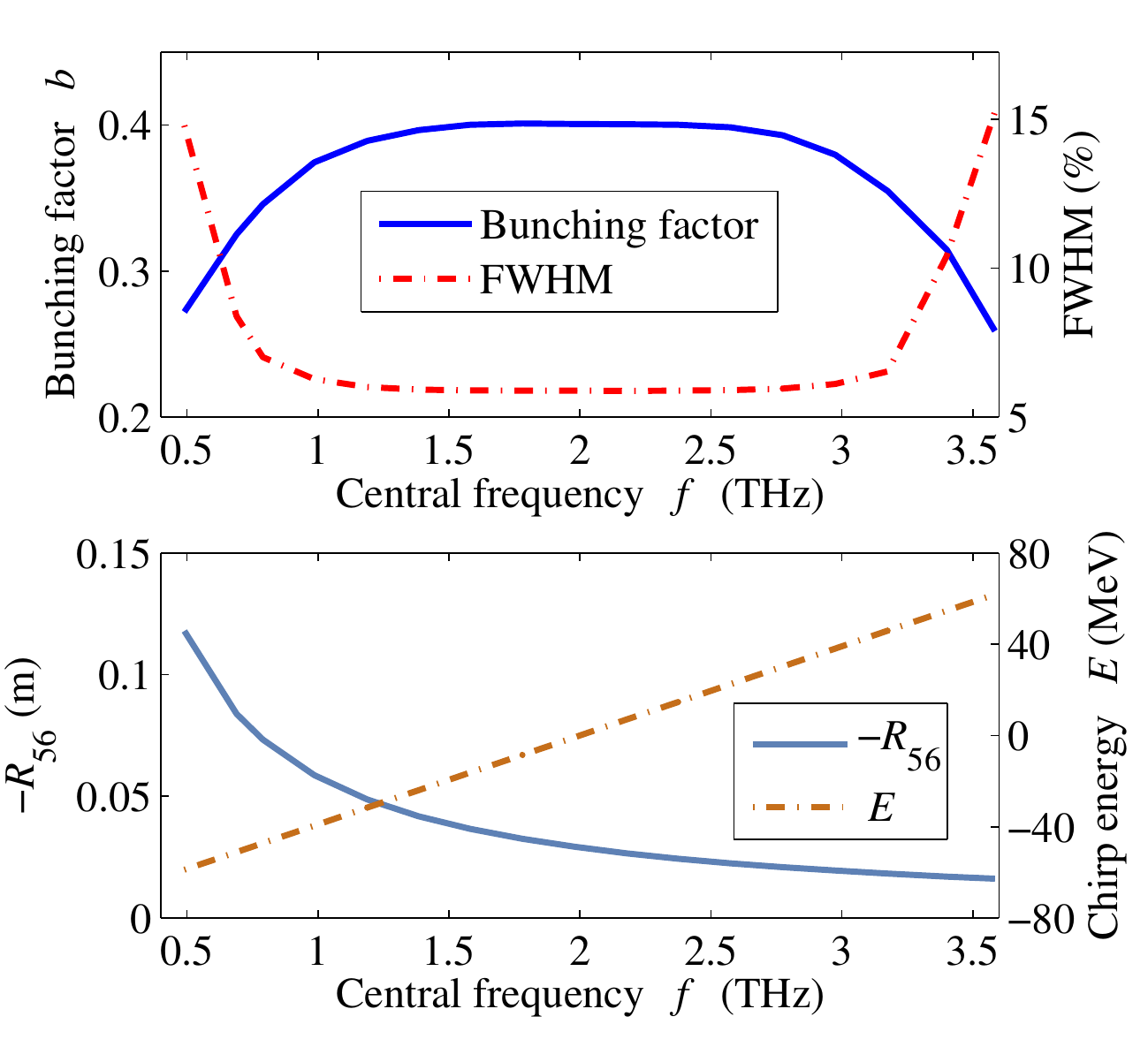}
   \caption{Bunching factor and the FWHM of the bunching spectra for different radiation frequency by matching the $R_{56}$ and chirp energy. The laser power is fixed at 1\,GW.}
   \label{scan_R56}
\end{figure}

In addition, we can also choose to vary the laser power and fix the $R_{56}$. Figure\,\ref{scan_P} presents the results of bunching factor, FWHM, required laser and chirp energy for different frequencies, which are similar with the results of varying $R_{56}$. In practice, we can vary both $R_{56}$ and laser power to optimize the bunching factor for different frequencies based on the availability of the beam line.
\begin{figure}[ht]
   \centering
   \includegraphics*[width=80mm]{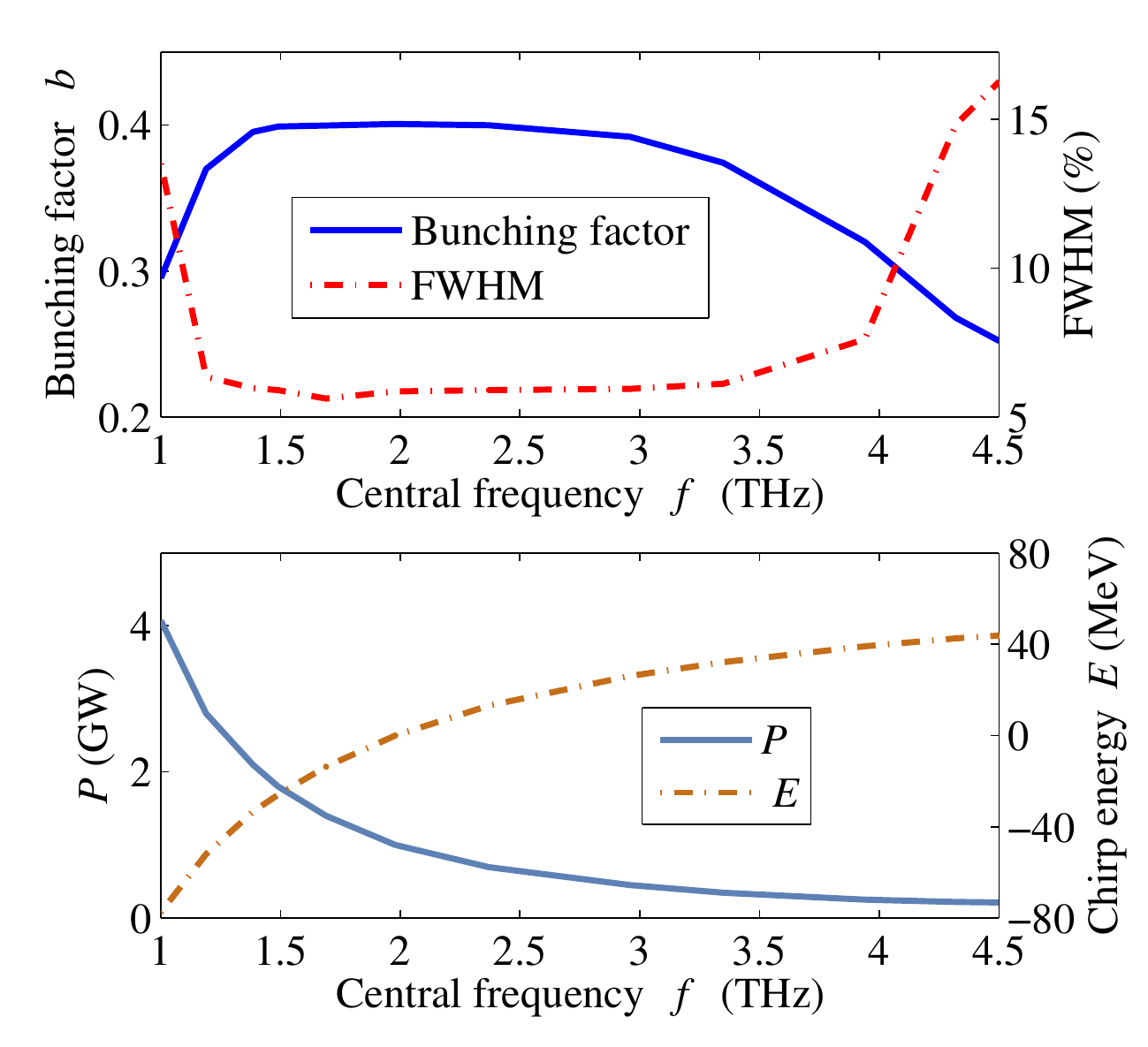}
   \caption{Bunching factor and the FWHM of the bunching spectra for different radiation frequency by matching the laser power and chirp energy. The $R_{56}$ is fixed at -29\,mm.}
   \label{scan_P}
\end{figure}

Lastly, we present some examples of 4\,THz density modulation simulations. As the frequency is increased by a factor of 2, the laser power is decreased to 0.26\,GW to meet the optimal condition with $R_{56}=-29$\,mm. By varying $R_{56}$, the bunching frequency can be varied between 2 to 6\,THz with a bunching factor larger than 0.3, as shown in Fig.\,\ref{X-band}. However, if we add a harmonic RF cavity that decelerates the beam at -180\,$^\circ$ before the chicane to compensate for the nonlinear effects during compression, the frequency range of the modulation can be extended significantly. In Fig.\,\ref{X-band}, we give the requirements for $R_{56}$, chirp energy, and especially the maximum energy gain of a 4th-harmonic (X-band) cavity to vary bunching frequency. With the help of chirp compensation from the X-band cavity (e.g. \,\cite{linearizer}), the bunching frequency range is extended from 1 to $\sim$\,8\,THz with the bunching factor keeping at $\sim$\,0.38.

\begin{figure}[ht]
   \centering
   \includegraphics*[width=80mm]{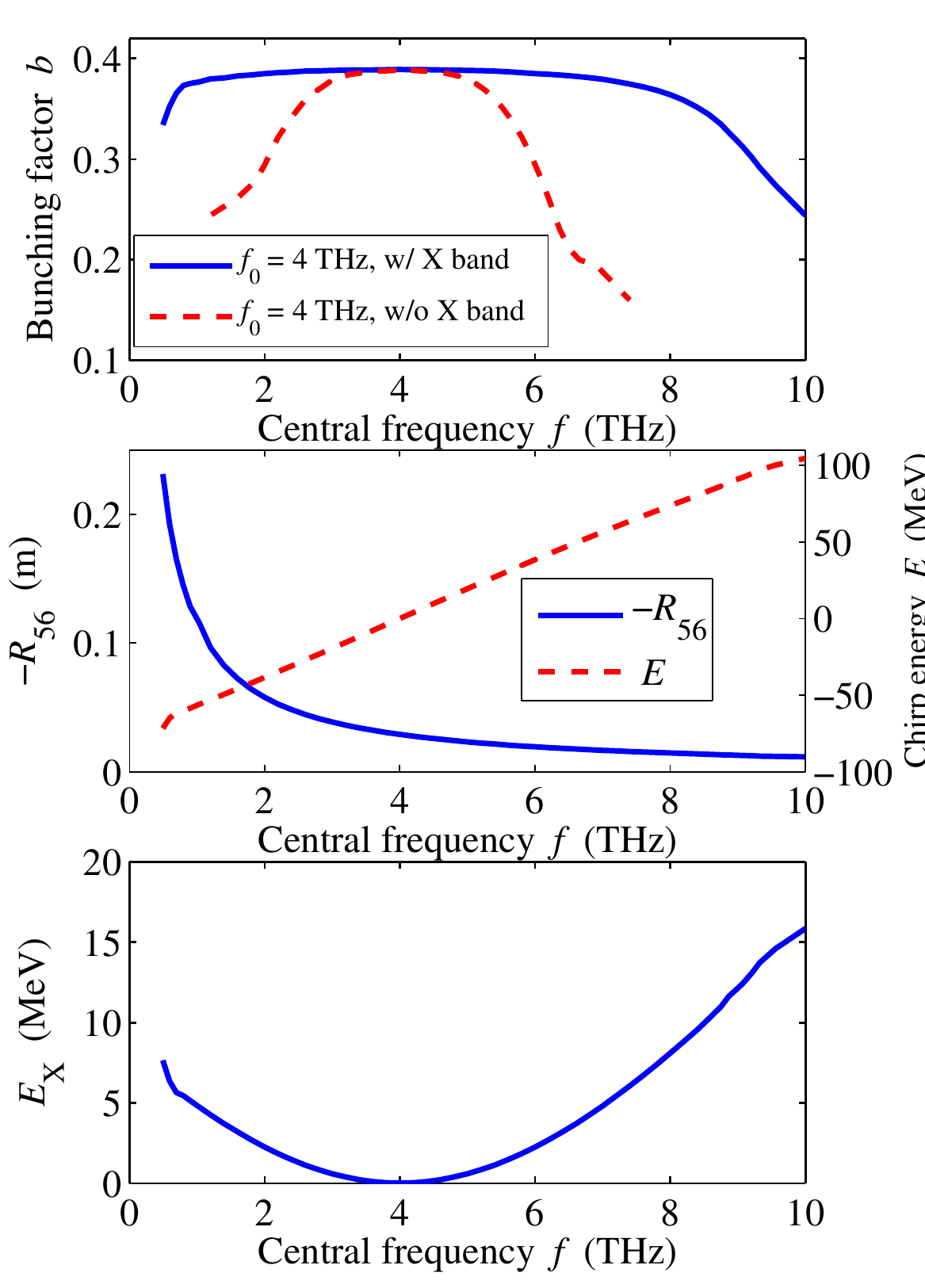}
   \caption{(top) Bunching factor for different radiation frequency w/ and w/o X-band cavity. (middle) The requirements for $R_{56}$ and chirp energy. (bottom) The requirement for energy gain of X-band cavity.}
   \label{X-band}
\end{figure}

Note that the bunching frequency control discussed above is to compress or stretch the electron beam. Alternatively we can also change the laser intensity modulation period directly to control the bunching frequency. As the increase of bunching frequency $k_1$, the $R_{56}$ of the chicane or (and) the laser power needs to be decreased to keep the optimal condition. Usually laser-induced energy spread should be much larger than the initial slice energy spread to introduce enough effective modulation depth, so reducing the $R_{56}$ of the chicane will be a better choice. Compared with the beam compression, in this method the nonlinear effects in the chicane will be smaller, which is helpful to maintain the density bunching. In addition, the number of sub-bunches in the train will become larger for higher modulation frequency, resulting in smaller bandwidth in the spectra.

\section{Discussions}\label{discussion}
The proposed method is of great advantages in the generations of tunable narrow-band THz radiation. The bunching factor averaged over the beam can be kept around 0.4 for a wide range of frequency, covering the THz gap, by varying the beam compression and (or) the laser modulation period. The proposed method is robust as it is performed on a very relativistic electron beam. Since there is no strong space charge force or beam loss during the process, the beam quality can be preserved for beam matching and focusing. Assuming that the beam is focused to a small spot and the frequency cut due to finite dimensions of the target, e.g., a metallic foil, can be neglected, it is possible to estimate the THz energy emitted by this beam as
\begin{align}
E_{\mbox{THz}}=N_e^2b^2\frac{dW_1}{d\omega}\Delta\omega\,,
\end{align}
where $b$ is the bunching factor, $N_e$ is the number of electrons for coherent emission, $\frac{dW_1}{d\omega}$ is the single particle spectral power which for transition radiation is relative relatively flat and equal to $\frac{2}{\pi}m_er_ec\ln\gamma$, with $m_e$ the rest electron mass and $r_e$ the classical electron radius. $\Delta\omega$ is the linewidth which is equal to 1 over the number of cycles in the pulse train. In the previous simulations, the beam charge is set at 0.5\,nC. Even though there is no strict constraint on beam charge and beam repetition rate. Using 2\,THz radiation as an example, the THz pulse energy emitted from 10-ps beam is shown in Fig.\,\ref{THz_energy} as a function of beam charge. When the charge is 2\,nC, the pulse energy is up to more than 40\,$\mu$J centered around 2\,THz frequency in 5\% bandwidth. The spectra brightness and THz field are both significantly improved. Finally, we note that using an undulator\,\cite{THz3} or a dielectric tube\,\cite{energy_modulation3,dielectric_tube1,dielectric_tube2} can further enhance the output THz energy to approach the mJ level.
\begin{figure}[ht]
   \centering
   \includegraphics*[width=80mm]{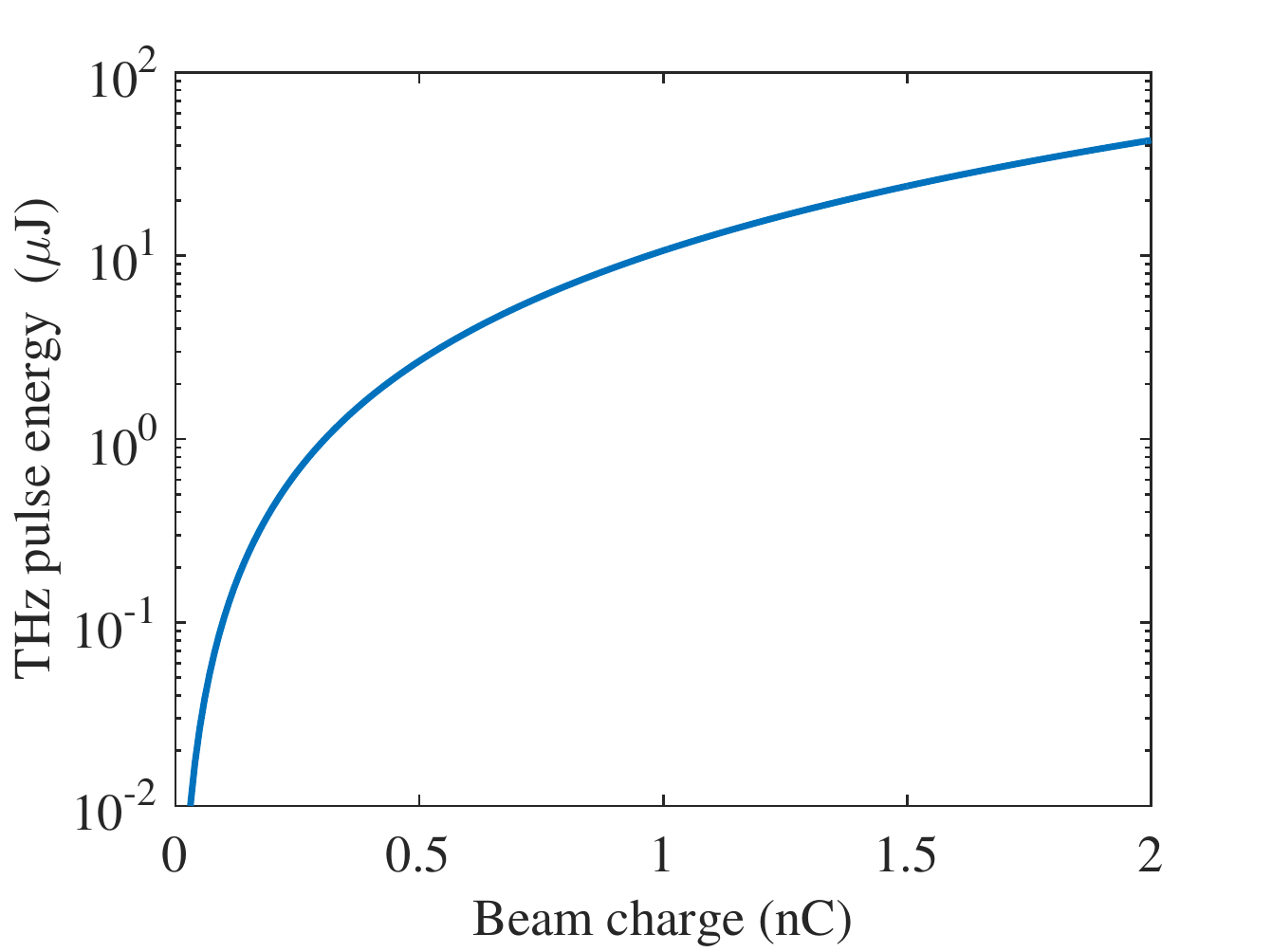}
   \caption{THz pulse energy around 2\,THz frequency in 5\% bandwidth emitted from 10-ps beam for different beam charge.}
   \label{THz_energy}
\end{figure}
 
In passing we note that our proposal does not require a linac with a photo-cathod RF gun. The method is also applicable for the electron beams from storage rings or thermal-cathode injectors with higher repetition rate. The main requirement of the method is the electron beam energy needs to be $\sim$\,100\,MeV to be resonant with an optical laser. However, for electron beams with lower energies, it is still possible to interact harmonically with the optical laser.

\section{Summary}\label{summary}
In this paper, we have proposed a method based on the slice energy spread modulation to generate strong density bunching in a relativistic electron beam, which can be used to produce intense narrow-band THz radiation. Theoretical analysis and simulations both show that with the help of double-horn slice energy spread distribution from the laser modulation, the bunching factor can reach up to 0.4 for modulation frequencies ranging from 1 to 10\,THz. We also found the optimal condition involving the bunching frequency, $R_{56}$ of the conversion section and the average slice energy spread to maximum the bunching factor. To implement this scheme in an existing x-ray FEL, very minimal hardware additions are required.  The use of the laser in our proposal makes the THz signal synchronize with the optical signal and has a stable waveform. The high spectra brightness of the radiation make it a powerful and promising method for many applications, such as THz pump\,\cite{THz_pump}, THz streaking\,\cite{THz_streak} and THz acceleration\,\cite{THz_acceleration}.

\section{Acknowledge}
We thank Enrico Allaria and Agostino Marinelli for very helpful discussions. This work was supported by the National Natural Science Foundation of China (NSFC Grants No. 11375097, No. 11435015 and No. 11475097) and  U.S. Department of Energy Contracts No. DE-AC02-76SF00515.


\begin{thebibliography}{} 
\bibitem{FLASH}
W. Ackermann  {\it et al.}, Nat. Photonics 1, 336 (2007).

\bibitem{LCLS}
P. Emma {\it et al.}, Nat. Photonics 4, 641 (2010).

\bibitem{SCALA}
T. Ishikawa {\it et al.}, Nat. Photonics 6, 540 (2012).

\bibitem{THz1}
G. L. Carr {\it et al.},  Nature 420(6912), 153-156 (2002).

\bibitem{THz2}
S. Bielawski {\it et al.}  Nature Physics, 4(5), 390-393 (2008).

\bibitem{plasma1}
I. Blumenfeld {\it et al.}  Nature, 445(7129), 741-744 (2007).

\bibitem{plasma2}
M. Litos {\it et al.}  Nature, 515(7525), 92-95 (2014).

\bibitem{dielectric}
C. Jing {\it et al.}, Phys. Rev. Lett. 98, 144801 (2007).

\bibitem{THz3}
A. Gover, Phys. Rev. ST Accel. Beams 8, 030701 (2005).

\bibitem{bunch_train1}
P. Chen, J. M. Dawson, R. W. Huff and T. Katsouleas, Phys. Rev. Lett. 54(7), 693 (1985).

\bibitem{bunch_train2}
R. D. Ruth, P. L. Morton, P. B. Wilson and  A. W. Chao, Part. Accel., 17, 171 (1984)  and SLAC-PUB-3374.

\bibitem{bunch_train3}
P. Schutt, T. Weiland and V. M. Tsakanov,  in {\it Proceedings of the Second All-Union Conference on New Methods of Charged Particle Acceleration} (Springer, New York, 1989).

\bibitem{bunch_train4}
J. G. Power, W. Gai, X. Sun and A. Kanareykin,  In Particle Accelerator Conference, 2001. PAC 2001. Proceedings of the 2001 (Vol. 1, pp. 114-116). IEEE.

\bibitem{multi-color}
E. Roussel {\it et al.} Phys. Rev. Lett. 115, 214801(2015).

\bibitem{sideband}
Z. Zhang {\it et al.} Phys. Rev. Accel. Beams 19, 050701 (2016).

\bibitem{exchange1}
P. Muggli {\it et al.}, Phys. Rev. Lett. 101, 054801 (2008).

\bibitem{exchange2}
Y. E. Sun {\it et al.}, Phys. Rev. Lett. 105, 234801 (2010).

\bibitem{laser_modulation1}
Y. Shen {\it et al.}, Phys. Rev. Lett. 107, 204801 (2011).

\bibitem{laser_modulation2}
Y. Li and K. J. Kim Appl. Phys. Lett. 92, 014101 (2008).

\bibitem{laser_modulation3}
P. Musumeci, R. K. Li, and A. Marinelli, Phys. Rev. Lett. 106, 184801 (2011).

\bibitem{laser_modulation4}
Z. Zhang {\it et al.}, Phys. Rev. Lett. 116, 184801 (2016).

\bibitem{energy_modulation1}
S. Antipov {\it et al.}, Phys. Rev. Lett. 108, 144801 (2012).

\bibitem{energy_modulation2}
K. Bane and G. Stupakov, Nucl. Instrum. Methods Phys. Res., Sect. A 677, 67 (2012).

\bibitem{energy_modulation3}
S. Antipov {\it et al.}, Phys. Rev. Lett. 111, 134802 (2013).

\bibitem{diff_frequency1}
D. Xiang and G. Stupakov, Phys. Rev. ST Accel. Beams 12, 080701 (2009).

\bibitem{diff_frequency2}
M. Dunning {\it et al.}, Phys. Rev. Lett. 109, 074801 (2012).

\bibitem{diff_frequency3}
Z. Wang, D. Huang, Q. Gu, Z. Zhao and D. Xiang, Phys. Rev. ST Accel. Beams 17, 090701 (2014).

\bibitem{laser_heater0}
E. L. Saldin, E. A. Schneidmiller and M.V. Yurkov, Nucl. Instrum. Methods Phys. Res., Sect. A 528, 355 (2004).

\bibitem{laser_heater1}
Z. Huang {\it et al.} Phys. Rev. ST Accel. Beams 7, 074401 (2004).

\bibitem{laser_heater2}
Z. Huang {\it et al.} Phys. Rev. ST Accel. Beams 13, 020703 (2010).

\bibitem{laser_heater3}
E. Ferrari {\it et al.} Phys. Rev. Lett. 112, 114802 (2014).

\bibitem{laser_heater4}
S. Spampinati {\it et al.} Phys. Rev. ST Accel. Beams 17, 120705 (2014).

\bibitem{laser_heater5}
J. Lee {\it et al.} Nucl. Instrum. Methods Phys. Res., Sect. A 843, 39 (2017).

\bibitem{chirped_beating}
A. S. Weling and D. H. Auston, J. Opt. Soc. Am. B 13, 2783 (1996).

\bibitem{BBO1}
J. Power and C. Jing, AIP Conf. Proc. 1086, 689 (2009)

\bibitem{BBO2}
P. Musumeci, J. T. Moody, C. M. Scoby, M. S. Gutierrez, M. Westfall, and R. K. Li, J. Appl. Phys. 108, 114513 (2010).

\bibitem{BBO3}
L. X. Yan {\it et al.} J. Plasma Phys. 78, 429 (2012).

\bibitem{elegant}
M. Borland, ELEGANT, Advanced Photon Source LS-287, 2000.

\bibitem{linearizer}
 P. Emma, “X-Band RF Harmonic Compensation for Linear Bunch Compression in the LCLS,” SLAC-TN-05-004 (2001).

\bibitem{dielectric_tube1}
A. M. Cook {\it et al.} Phys. Rev. Lett. 103, 095003 (2009).

\bibitem{dielectric_tube2}
G. Andonian, {\it et al.} Appl. Phys. Lett. 98, 202901 (2011).

\bibitem{THz_pump}
M. Liu {\it et al.} {\it Nature} 487, 345–348 (2012).

\bibitem{THz_streak}
U. Fr{\"u}hling {\it et al.} {\it Nature Photonics} 3, 523-528 (2009).

\bibitem{THz_acceleration}
E. A. Nanni {\it et al.} {\it Nature Communications} 6, 8486 (2015).

\end{thebibliography}
\end{document}